\documentclass[twocolumn,superscriptaddress,showpacs,preprintnumbers,amsmath,amssymb,aps,prb]{revtex4}
\usepackage{graphicx}
\usepackage{dcolumn}
\usepackage{bm}
\usepackage{color}

\begin{document}
\title{Comments on the elastic properties in solid solutions of silver halides}
\author{E. S. Skordas}
\affiliation{Section of Solid State Physics and Solid Earth Physics
Institute, Physics Department, University of Athens,
Panepistimiopolis, Zografos 157 84, Athens, Greece}

\begin{abstract}
Recently first principles microscopic calculations, using the generalized gradient approximation, appeared for the solid mixed system AgCl${_x}$Br$_{1-x}$ at various compositions. Here, we suggest a model that can estimate the compressibility of the mixed crystals in terms of the compressibilities of the end members alone. This model makes use of a single parameter, i.e., the compressibility of a defect volume, when considering the volume variation produced by the addition of a "foreign molecule" to a host crystal as a defect volume. 
\end{abstract} 
 \maketitle

\section{Introduction}

The alkali halides are the most simple ionic solids and the study of their elastic properties has revealed useful information about interatomic potentials and forces. Their description in terms of the well known Born model is satisfactory. To the contrary, the silver halides exhibit unusual properties compared to the alkali halides, such as appreciably lower melting point and ionic conductivity larger by several orders of magnitude. This is one of the basic reasons that the study of physical properties of silver halides have recently attracted an intense interest. Examples of these recent studies are: First, calculations from first principles\cite{ref1} of the structural, electronic and thermodynamic properties of silver halides as well as of their rock-salt AgCl$_x$Br$_{1-x}$  alloys by applying the full potential linearized augmented plane wave method have recently appeared. For the alloys, the effect of composition on lattice constants, bulk modulus, cohesive energy, bond ionicity, band gap and effective mass was also investigated. Second, measurements of the temperature dependence of their elastic constants by means of the resonant ultrasound spectrospopy method have been recently reported,\cite{ref2}  which overcome the difficulty in determining elastic constants of small samples by pulse echo methods because of the short travelling time of pulse echoes. Third, the phonons and elastic constants AgCl and AgBr under pressure have been extensively studied\cite{ref3}  by means of the pseudopotential plane-wave method within density functional theory. Finally, the elastic properties and the lattice dynamics of AgBr$_{1-x}$Cl$_x$ have been studied as a function of composition ($x$) using the density functional perturbation theory and employing the virtual-crystal approximation by Bouamama et al.\cite{ref4} They used both the local density approximation (LDA) and the generalized gradient approximation (GGA) and found that GGA gives more appropriate elastic constants in agreement with the experimental values.\cite{ref5} 

The structure of the intrinsic point defects in solid silver halides, and in particular AgCl and AgBr, is of both fundamental and industrial importance. Due to their unique defect properties, the silver halides play a major role on the photographic industry. The dominant defects in the silver halides are Frenkel defect pairs, consisting of interstitial cations and corresponding vacancies (e.g., see Ref. 
\onlinecite{ref6} and references therein). It is a well known experimental fact that in AgCl and AgBr their electrical conductivity plot $\ln(\sigma T)$ versus $1/T$ exhibits a strong upwards curvature in the high temperature ($T$). \cite{ref7,ref8,ref9} This effect has been attributed mainly to the non-linear decrease of the defect Gibbs formation energy\cite{ref9} (while in most cases this decrease is linear\cite{ref10}) upon increasing the temperature, which is accompanied by a simultaneous increase of the defect formation enthalpy and entropy versus the temperature. Note that this entropy differs essentially from the dynamic entropy defined recently in natural time.\cite{ref11}

It is the basic objective of this paper to show the following interesting fact. The recent tedious GGA calculations of Bouamama et al.\cite{ref4}  in AgBr$_{1-x}$Cl$_x$ lead to results that are in striking agreement with those deduced from a simple thermodynamical model that interconnects the defect formation and/or migration Gibbs energies with bulk properties. This thermodynamical model –termed $cB\Omega$ model\cite{ref12,ref13,ref14,ref15,ref16}  (see below)- that has been employed for the estimation of the compressibility of the mixed alkali halide crystals (the dielectric constants of which have been studied in Ref. \onlinecite{ref17}). 

\section{The method to describe the compressibility of a mixed system.}

Let us first explain how the compressibility  $\kappa$  of a mixed system A$_x$B$_{1-x}$ can be determined in terms of the compressibilities of the two end members A and B by following Refs. \onlinecite{ref6} and \onlinecite{ref18}. We call the two end members A and B as pure components ``1''  and ``2'', respectively and label $v_1$  the volume per ``molecule’’ of the pure component ``1''  (assumed to be the major component in the aforementioned mixed system  A$_x$B$_{1-x}$ ),$v_2$ the volume per ``molecule'' of the pure component ``2''. 
Furthermore, let denote $V_1$ and $V_2$ the corresponding molar volumes, i.e. $V_1= N v_1$ and 
$V_2=  N v_2$ (where $N$ stands for Avogadro’s number) and assume that $v_1<v_2$ . 
Defining a ``defect volume''\cite{ref6,ref18} $ v^d$  as the increase of the volume $V_1$, if one 
“molecule'' of type ``1'' is replaced by one “molecule'' of type ``2'', it is evident that the addition of one “molecule'' of type ``2'' to a crystal containing $N$ ``molecule'' of type ``1''  will increase its volume by $v^d+v_1$  (see Chapter 12 of Ref. \onlinecite{ref6} as well as Ref. \onlinecite{ref18}). Assuming that $v^d$  is independent of composition, the volume $V_{N+n}$  of a crystal containing $N$ ``molecule'' of type ``1'' and n “molecules” of type ``2''  is given by:
\begin{equation}
V_{N+n}=Nv_1+n(v^d+v_1)  \phantom{x} {\rm or} \phantom{x} [1+(n/N)]V_1+nv^d 
\end{equation} 
The compressibility $\kappa$  of the mixed system can be found by differentiating (1) with respect to pressure which finally gives:  
\begin{equation}
\kappa V_{N+n}=\kappa_1 V_1 +(n/N) [\kappa^d N v^d+\kappa_1 V_1]
\end{equation} 
where $\kappa^d$ denotes the compressibility of the volume $v^d$ , defined as $\kappa^d=-(1/v^d)(dv^d/dP)_T$.
The ``defect-volume'' $v^d$ can be approximated from:\cite{ref6,ref18}
\begin{equation}
v^d=(V_2-V_1)/N  \phantom{x} {\rm or} \phantom{x} v^d=v_2-v_1
\end{equation} 
In view of the fact that $V_{N+n}$  can be calculated from (1) –when taken into account (3)- the compressibility $\kappa$ can be determined from (2) provided that an estimation of $v^d$ can be envisaged. Towards this goal, we employ the aforementioned thermodynamical model for the formation and migration of the defects in solids. This model, which will be shortly described below, has been successfully applied to various categories of solids\cite{ref12,ref13,ref14,ref15,ref16,ref19,ref20}  as well as to complex ionic materials under uniaxial stress that emit electric signals before fracture, thus explaining the signals detected\cite{ref21,ref22,EPL12} before major earthquakes. 

\begin{figure}
\includegraphics[scale=0.45]{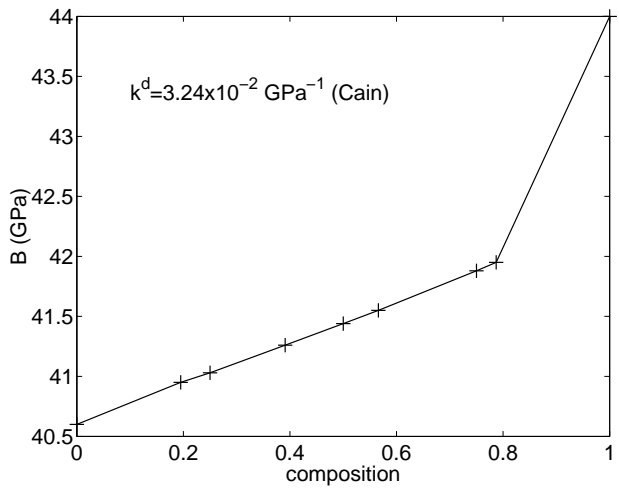}
\caption{Plot of the quantity $\kappa V_{N+n}$ versus $n/N$ in the mixed system AgBr$_{1-x}$Cl$_x$ when using the experimental values\cite{ref5} of the compressibility $\kappa$  at various compositions $x$. }\label{f1}
\end{figure}

\begin{figure}
\includegraphics[scale=0.45]{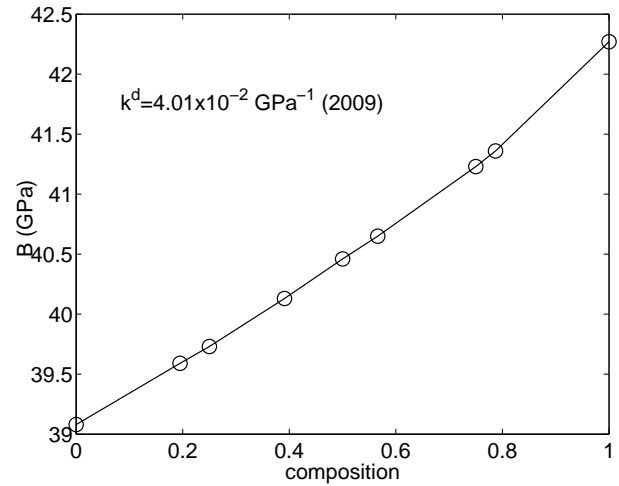}
\caption{Plot of the quantity $\kappa V_{N+n}$ versus $n/N$ in the mixed system AgCl$_x$Br$_{1-x}$  when using the theoretical $\kappa$-values deduced in Ref. \onlinecite{ref4} –by means of GGA- at various compositions $x$.}\label{f2}
\end{figure}

Within the frame of this thermodynamical model (usually called $cB\Omega$ model as mentioned), the defect Gibbs energy $g^i$ is interrelated with the bulk properties of the solid through the relation $g^i=c^i B\Omega$  where $B$ stands for the isothermal bulk modulus, $\Omega$  the mean volume per atom and $c^i$ is dimensionless quantity. The superscript $i$ refers to the type of the defect process under consideration ($i$ = defect formation, migration, etc.). By differentiating this relation in respect to pressure $P$, we find that the defect volume $v^i [=(dg^i/dP)_T]$ , and the compressibility $\kappa^{d,i}$, defined by  $\kappa^{d,i} \equiv -(1/v^i)(dv^i/dP)_T$, is obtained:\cite{ref6}
\begin{equation}
\kappa^{d,i}=(1/B)-(d^2B/dP^2)/[(dB/dP)_T-1]
\end{equation}

Since, this relation reveals that the compressibility $\kappa^{d,i}$, does not change upon changing the type $i$ of the defect process, it is reasonable to assume that (4) holds also for the compressibility $\kappa^d$ of the ``defect’’ volume $v^d$  in (2) and hence we have 
\begin{equation}
\kappa^{d}=(1/B_1)-(d^2B_1/dP^2)/[(dB_1/dP)_T-1]
\end{equation} 
where the subscript “1” in the quantities at the right side denotes that they refer to the pure component (1).  

\section{Results for the compressibility in the system $AgBr_{1-x}Cl_{x}$}

Let us now apply the procedure explained in the previous section to the mixed system
AgBr$_{1-x}$Cl$_x$. We take as starting material AgBr ``1''  ( $V_1 $=28.996 cm$^3$/mole) and by
considering that for the pure AgCl ``2''  the volume is $V_2 $=25.731 cm$^3$/mole, one finds
$Nv^d=V_2=V_1$=-3.265cm$^3$.

We now consider the adiabatic compressibility values measured in Ref.\onlinecite{ref5} and plot
in Fig. 1, for various compositions $x$, the quantity $\kappa V_{N+n}$ versus $n/N$. 
We find that it
is approximately a straight line the slope of which, according to (2), is
$\kappa^d (Nv^d) + \kappa_1 V_1 = 6.4 \times 10^{-1}$ cm$^3$GPa$^{-1}$. 
By inserting the aforementioned $v^d$ -value, we
get $\kappa^d=3.95 \times 10^{2}$ GPa$^{-1}$. 
By the same token we now repeat the same procedure, but
now using –instead of the experimental values- the calculated compressibility values
by means of GGA reported in Ref.\onlinecite{ref4}. We now find the plot given in Fig. 2 from the
slope of which we get $\kappa^d \approx 4.0 \times  10^{-2}$ GPa$^{-1}$ in nice agreement with 
the $\kappa^d$-value
obtained from the experimental $\kappa$-values in Fig. 1. If we alternatively use the
calculated compressibility values by means of LDA reported in Ref. \onlinecite{ref4}, we finally
obtain the value $\kappa^d \approx 5.0 \times  10^{-2}$GPa$^{-1}$  which markedly differs from the value
$\kappa^d= 3.95 \times 10^{-2}$GPa$^{-1}$ deduced from the experimental values in Fig. 1. The latter
difference is consistent with the findings in Ref.\onlinecite{ref4} according to which a better
agreement between the calculated elastic constants with composition and the
experimental data is found within the GGA.

We finally proceed to the calculation of $\kappa^d$  on the basis of (5), by using the elastic
data under pressure reported by Loje and Schuele.\cite{ref23} They found that isothermal bulk
modulus of AgBr depends on pressure P
through: $B(p)=377.7+7.49P-0.0287P^2/2$, where $B$ and $P$ are in kilobars thus
$(dB/dP)_T=7.49$  and $(d^2B/dP^2)_T=-0.0287$Kbar$^{-1}$.

By inserting these values into (5) we find $\kappa^d=(7.0\pm 3.0)\times 10 ^{-2}$GPa$^{-1}$, if the
experimental errors reported in Ref. \onlinecite{ref23} are envisaged. (The large uncertainty of more
than 40\% in the estimation of the$\kappa^d$ -value comes mainly from the large experimental
uncertainty in the determination of the quantity $(d^2B/dP^2)_T$  in Ref.\onlinecite{ref23}  when fitting
the elastic data of AgBr through a second order Murnaghan equation of state; it is the
nonzero value of this quantity which leads to the conclusion that the defect
compressibility $\kappa^d$ significantly exceeds the bulk compressibility $1/B_1$ , e.g., see Refs.
\onlinecite{ref24,ref25}). Quite interestingly, the lower bound of $\kappa^d$ ($4.0 \times 10^{-2}$ GPa$^{-1}$) agrees with the
$\kappa^d$ - value resulted from Fig. 2, i.e., when considering the $\kappa$-values calculated in Ref. \onlinecite{ref4}
by means of GGA. This agreement is remarkable if we consider that the $cB\Omega$-model
deduced the $\kappa^d$-value on the basis of (5) by making use of the elastic data of AgBr
alone. It is believed that the current procedure based on Eq. (5) can be also applied to
other systems instead of silver halides provided that elastic data of the pure
constituents are available which have led to the determination of the quantity
$(d^2B_1/dP^2)_T$   involved in Eq. (5).

\section{Main Conclusions}
Here, we assume that the volume variation produced by the addition of a ``foreign
molecule'' to a host crystal can be considered as a defect volume. Then we show that
when using as a single parameter the compressibility $\kappa^d$ of the defect volume, we can
estimate the compressibility $\kappa$ of the mixed system AgBr$_{1-x}$Cl$_x$ in terms of the
compressibilities of the end members. Interestingly, the $\kappa^d$-value obtained from the
analysis of the experimental data at various compositions is found (Fig. 1) to agree
reasonably well with the one deduced from the analysis of the theoretical values (Fig.
2) derived by Bouamama et al.\cite{ref4}  within the GGA approximation. Moreover, we found
that these $\kappa^d$-values agree with the corresponding $\kappa^d$-value resulted from the
thermodynamical $cB\Omega$-model on the basis of the elastic data of AgBr alone.


\begin{thebibliography}{26}
\expandafter\ifx\csname natexlab\endcsname\relax\def\natexlab#1{#1}\fi
\expandafter\ifx\csname bibnamefont\endcsname\relax
  \def\bibnamefont#1{#1}\fi
\expandafter\ifx\csname bibfnamefont\endcsname\relax
  \def\bibfnamefont#1{#1}\fi
\expandafter\ifx\csname citenamefont\endcsname\relax
  \def\citenamefont#1{#1}\fi
\expandafter\ifx\csname url\endcsname\relax
  \def\url#1{\texttt{#1}}\fi
\expandafter\ifx\csname urlprefix\endcsname\relax\def\urlprefix{URL }\fi
\providecommand{\bibinfo}[2]{#2}
\providecommand{\eprint}[2][]{\url{#2}}

\bibitem[{\citenamefont{Amrani et~al.}(2007)\citenamefont{Amrani, Hassan, and
  Zoaeter}}]{ref1}
\bibinfo{author}{\bibfnamefont{B.}~\bibnamefont{Amrani}},
  \bibinfo{author}{\bibfnamefont{F.~E.~H.} \bibnamefont{Hassan}},
  \bibnamefont{and} \bibinfo{author}{\bibfnamefont{M.}~\bibnamefont{Zoaeter}},
  \bibinfo{journal}{Physica B: Condensed Matter}
  \textbf{\bibinfo{volume}{396}}, \bibinfo{pages}{192 } (\bibinfo{year}{2007}).

\bibitem[{\citenamefont{Endou et~al.}(2009)\citenamefont{Endou, Michihiro,
  Itsuki, Nakamura, and Ohno}}]{ref2}
\bibinfo{author}{\bibfnamefont{S.}~\bibnamefont{Endou}},
  \bibinfo{author}{\bibfnamefont{Y.}~\bibnamefont{Michihiro}},
  \bibinfo{author}{\bibfnamefont{K.}~\bibnamefont{Itsuki}},
  \bibinfo{author}{\bibfnamefont{K.}~\bibnamefont{Nakamura}}, \bibnamefont{and}
  \bibinfo{author}{\bibfnamefont{T.}~\bibnamefont{Ohno}},
  \bibinfo{journal}{Solid State Ionics} \textbf{\bibinfo{volume}{180}},
  \bibinfo{pages}{488 } (\bibinfo{year}{2009}).

\bibitem[{\citenamefont{Li et~al.}(2006)\citenamefont{Li, Zhang, Cui, Ma, Zou,
  and Klug}}]{ref3}
\bibinfo{author}{\bibfnamefont{Y.}~\bibnamefont{Li}},
  \bibinfo{author}{\bibfnamefont{L.}~\bibnamefont{Zhang}},
  \bibinfo{author}{\bibfnamefont{T.}~\bibnamefont{Cui}},
  \bibinfo{author}{\bibfnamefont{Y.}~\bibnamefont{Ma}},
  \bibinfo{author}{\bibfnamefont{G.}~\bibnamefont{Zou}}, \bibnamefont{and}
  \bibinfo{author}{\bibfnamefont{D.~D.} \bibnamefont{Klug}},
  \bibinfo{journal}{Phys. Rev. B} \textbf{\bibinfo{volume}{74}},
  \bibinfo{pages}{054102} (\bibinfo{year}{2006}).

\bibitem[{\citenamefont{Bouamama et~al.}(2009)\citenamefont{Bouamama, Djemia,
  Daoud, and Ch\'{e}rif}}]{ref4}
\bibinfo{author}{\bibfnamefont{K.}~\bibnamefont{Bouamama}},
  \bibinfo{author}{\bibfnamefont{P.}~\bibnamefont{Djemia}},
  \bibinfo{author}{\bibfnamefont{K.}~\bibnamefont{Daoud}}, \bibnamefont{and}
  \bibinfo{author}{\bibfnamefont{S.}~\bibnamefont{Ch\'{e}rif}},
  \bibinfo{journal}{Computational Materials Science}
  \textbf{\bibinfo{volume}{47}}, \bibinfo{pages}{308 } (\bibinfo{year}{2009}).

\bibitem[{\citenamefont{Cain}(1977)}]{ref5}
\bibinfo{author}{\bibfnamefont{L.}~\bibnamefont{Cain}},
  \bibinfo{journal}{Journal of Physics and Chemistry of Solids}
  \textbf{\bibinfo{volume}{38}}, \bibinfo{pages}{73 } (\bibinfo{year}{1977}).

\bibitem[{\citenamefont{Varotsos and Alexopoulos}(1986)}]{ref6}
\bibinfo{author}{\bibfnamefont{P.}~\bibnamefont{Varotsos}} \bibnamefont{and}
  \bibinfo{author}{\bibfnamefont{K.}~\bibnamefont{Alexopoulos}},
  \emph{\bibinfo{title}{Thermodynamics of Point Defects and their Relation with
  Bulk Properties}} (\bibinfo{publisher}{North Holland},
  \bibinfo{address}{Amsterdam}, \bibinfo{year}{1986}).

\bibitem[{\citenamefont{Batra and Slifkin}(1977)}]{ref7}
\bibinfo{author}{\bibfnamefont{A.}~\bibnamefont{Batra}} \bibnamefont{and}
  \bibinfo{author}{\bibfnamefont{L.}~\bibnamefont{Slifkin}},
  \bibinfo{journal}{Journal of Physics and Chemistry of Solids}
  \textbf{\bibinfo{volume}{38}}, \bibinfo{pages}{687 } (\bibinfo{year}{1977}).

\bibitem[{\citenamefont{Batra et~al.}(1980)\citenamefont{Batra, Hernandez, and
  Slifkin}}]{ref8}
\bibinfo{author}{\bibfnamefont{A.~P.} \bibnamefont{Batra}},
  \bibinfo{author}{\bibfnamefont{J.~P.} \bibnamefont{Hernandez}},
  \bibnamefont{and} \bibinfo{author}{\bibfnamefont{L.~M.}
  \bibnamefont{Slifkin}}, \bibinfo{journal}{Phys. Rev. B}
  \textbf{\bibinfo{volume}{22}}, \bibinfo{pages}{734} (\bibinfo{year}{1980}).

\bibitem[{\citenamefont{Varotsos and Alexopoulos}(1978{\natexlab{a}})}]{ref9}
\bibinfo{author}{\bibfnamefont{P.}~\bibnamefont{Varotsos}} \bibnamefont{and}
  \bibinfo{author}{\bibfnamefont{K.}~\bibnamefont{Alexopoulos}},
  \bibinfo{journal}{J. Phys. Chem. Solids} \textbf{\bibinfo{volume}{39}},
  \bibinfo{pages}{759 } (\bibinfo{year}{1978}{\natexlab{a}}).

\bibitem[{\citenamefont{Varotsos and Miliotis}(1974)}]{ref10}
\bibinfo{author}{\bibfnamefont{P.}~\bibnamefont{Varotsos}} \bibnamefont{and}
  \bibinfo{author}{\bibfnamefont{D.}~\bibnamefont{Miliotis}},
  \bibinfo{journal}{Journal of Physics and Chemistry of Solids}
  \textbf{\bibinfo{volume}{35}}, \bibinfo{pages}{927 } (\bibinfo{year}{1974}).

\bibitem[{\citenamefont{Varotsos et~al.}(2011)\citenamefont{Varotsos, Sarlis,
  Skordas, Uyeda, and Kamogawa}}]{ref11}
\bibinfo{author}{\bibfnamefont{P.}~\bibnamefont{Varotsos}},
  \bibinfo{author}{\bibfnamefont{N.~V.} \bibnamefont{Sarlis}},
  \bibinfo{author}{\bibfnamefont{E.~S.} \bibnamefont{Skordas}},
  \bibinfo{author}{\bibfnamefont{S.}~\bibnamefont{Uyeda}}, \bibnamefont{and}
  \bibinfo{author}{\bibfnamefont{M.}~\bibnamefont{Kamogawa}},
  \bibinfo{journal}{Proc. Natl. Acad. Sci. USA} \textbf{\bibinfo{volume}{108}},
  \bibinfo{pages}{11361} (\bibinfo{year}{2011}).

\bibitem[{\citenamefont{Varotsos and Alexopoulos}(1980{\natexlab{a}})}]{ref12}
\bibinfo{author}{\bibfnamefont{P.}~\bibnamefont{Varotsos}} \bibnamefont{and}
  \bibinfo{author}{\bibfnamefont{K.}~\bibnamefont{Alexopoulos}},
  \bibinfo{journal}{Phys. Rev. B} \textbf{\bibinfo{volume}{21}},
  \bibinfo{pages}{4898} (\bibinfo{year}{1980}{\natexlab{a}}).

\bibitem[{\citenamefont{Varotsos}(1980)}]{ref13}
\bibinfo{author}{\bibfnamefont{P.}~\bibnamefont{Varotsos}},
  \bibinfo{journal}{physica status solidi (b)} \textbf{\bibinfo{volume}{100}},
  \bibinfo{pages}{K133} (\bibinfo{year}{1980}).

\bibitem[{\citenamefont{Varotsos and Alexopoulos}(1982)}]{ref14}
\bibinfo{author}{\bibfnamefont{P.}~\bibnamefont{Varotsos}} \bibnamefont{and}
  \bibinfo{author}{\bibfnamefont{K.}~\bibnamefont{Alexopoulos}},
  \bibinfo{journal}{physica status solidi (b)} \textbf{\bibinfo{volume}{110}},
  \bibinfo{pages}{9} (\bibinfo{year}{1982}).

\bibitem[{\citenamefont{Varotsos et~al.}(1982)\citenamefont{Varotsos,
  Alexopoulos, and Nomicos}}]{ref15}
\bibinfo{author}{\bibfnamefont{P.}~\bibnamefont{Varotsos}},
  \bibinfo{author}{\bibfnamefont{K.}~\bibnamefont{Alexopoulos}},
  \bibnamefont{and} \bibinfo{author}{\bibfnamefont{K.}~\bibnamefont{Nomicos}},
  \bibinfo{journal}{physica status solidi (b)} \textbf{\bibinfo{volume}{111}},
  \bibinfo{pages}{581} (\bibinfo{year}{1982}).

\bibitem[{\citenamefont{Varotsos}(2008)}]{ref16}
\bibinfo{author}{\bibfnamefont{P.}~\bibnamefont{Varotsos}},
  \bibinfo{journal}{Solid State Ionics} \textbf{\bibinfo{volume}{179}},
  \bibinfo{pages}{438 } (\bibinfo{year}{2008}).

\bibitem[{\citenamefont{Varotsos and Alexopoulos}(1978{\natexlab{b}})}]{ref17}
\bibinfo{author}{\bibfnamefont{P.}~\bibnamefont{Varotsos}} \bibnamefont{and}
  \bibinfo{author}{\bibfnamefont{K.}~\bibnamefont{Alexopoulos}},
  \bibinfo{journal}{physica status solidi (a)} \textbf{\bibinfo{volume}{47}},
  \bibinfo{pages}{K133} (\bibinfo{year}{1978}{\natexlab{b}}).

\bibitem[{\citenamefont{Varotsos and Alexopoulos}(1980{\natexlab{b}})}]{ref18}
\bibinfo{author}{\bibfnamefont{P.}~\bibnamefont{Varotsos}} \bibnamefont{and}
  \bibinfo{author}{\bibfnamefont{K.}~\bibnamefont{Alexopoulos}},
  \bibinfo{journal}{Journal of Physics and Chemistry of Solids}
  \textbf{\bibinfo{volume}{41}}, \bibinfo{pages}{1291 }
  (\bibinfo{year}{1980}{\natexlab{b}}).

\bibitem[{\citenamefont{Varotsos and Alexopoulos}(1980{\natexlab{c}})}]{ref19}
\bibinfo{author}{\bibfnamefont{P.}~\bibnamefont{Varotsos}} \bibnamefont{and}
  \bibinfo{author}{\bibfnamefont{K.}~\bibnamefont{Alexopoulos}},
  \bibinfo{journal}{J. Phys. Chem. Sol.} \textbf{\bibinfo{volume}{41}},
  \bibinfo{pages}{443} (\bibinfo{year}{1980}{\natexlab{c}}).

\bibitem[{\citenamefont{Varotsos and Alexopoulos}(1981)}]{ref20}
\bibinfo{author}{\bibfnamefont{P.}~\bibnamefont{Varotsos}} \bibnamefont{and}
  \bibinfo{author}{\bibfnamefont{K.}~\bibnamefont{Alexopoulos}},
  \bibinfo{journal}{J. Phys. Chem. Sol.} \textbf{\bibinfo{volume}{42}},
  \bibinfo{pages}{409} (\bibinfo{year}{1981}).

\bibitem[{\citenamefont{Varotsos et~al.}(2002)\citenamefont{Varotsos, Sarlis,
  and Skordas}}]{ref21}
\bibinfo{author}{\bibfnamefont{P.~A.} \bibnamefont{Varotsos}},
  \bibinfo{author}{\bibfnamefont{N.~V.} \bibnamefont{Sarlis}},
  \bibnamefont{and} \bibinfo{author}{\bibfnamefont{E.~S.}
  \bibnamefont{Skordas}}, \bibinfo{journal}{Acta Geophysica Polonica}
  \textbf{\bibinfo{volume}{50}}, \bibinfo{pages}{337} (\bibinfo{year}{2002}).

\bibitem[{\citenamefont{Varotsos et~al.}(1996)\citenamefont{Varotsos, Eftaxias,
  Lazaridou, Nomicos, Sarlis, Bogris, Makris, Antonopoulos, and
  Kopanas}}]{ref22}
\bibinfo{author}{\bibfnamefont{P.}~\bibnamefont{Varotsos}},
  \bibinfo{author}{\bibfnamefont{K.}~\bibnamefont{Eftaxias}},
  \bibinfo{author}{\bibfnamefont{M.}~\bibnamefont{Lazaridou}},
  \bibinfo{author}{\bibfnamefont{K.}~\bibnamefont{Nomicos}},
  \bibinfo{author}{\bibfnamefont{N.}~\bibnamefont{Sarlis}},
  \bibinfo{author}{\bibfnamefont{N.}~\bibnamefont{Bogris}},
  \bibinfo{author}{\bibfnamefont{J.}~\bibnamefont{Makris}},
  \bibinfo{author}{\bibfnamefont{G.}~\bibnamefont{Antonopoulos}},
  \bibnamefont{and} \bibinfo{author}{\bibfnamefont{J.}~\bibnamefont{Kopanas}},
  \bibinfo{journal}{Acta Geophysica Polonica} \textbf{\bibinfo{volume}{44}},
  \bibinfo{pages}{301} (\bibinfo{year}{1996}).

\bibitem[{\citenamefont{Varotsos et~al.}(2012)\citenamefont{Varotsos, Sarlis,
  and Skordas}}]{EPL12}
\bibinfo{author}{\bibfnamefont{P.}~\bibnamefont{Varotsos}},
  \bibinfo{author}{\bibfnamefont{N.}~\bibnamefont{Sarlis}}, \bibnamefont{and}
  \bibinfo{author}{\bibfnamefont{E.}~\bibnamefont{Skordas}},
  \bibinfo{journal}{EPL(Europhysics Letters)} \textbf{\bibinfo{volume}{99}},
  \bibinfo{eid}{59001} (\bibinfo{year}{2012}).

\bibitem[{\citenamefont{Loje and Schuele}(1970)}]{ref23}
\bibinfo{author}{\bibfnamefont{K.}~\bibnamefont{Loje}} \bibnamefont{and}
  \bibinfo{author}{\bibfnamefont{D.}~\bibnamefont{Schuele}},
  \bibinfo{journal}{Journal of Physics and Chemistry of Solids}
  \textbf{\bibinfo{volume}{31}}, \bibinfo{pages}{2051 } (\bibinfo{year}{1970}).

\bibitem[{\citenamefont{Varotsos and Ludwig}(1978)}]{ref24}
\bibinfo{author}{\bibfnamefont{P.}~\bibnamefont{Varotsos}} \bibnamefont{and}
  \bibinfo{author}{\bibfnamefont{W.}~\bibnamefont{Ludwig}},
  \bibinfo{journal}{Journal of Physics C: Solid State Physics}
  \textbf{\bibinfo{volume}{11}}, \bibinfo{pages}{L305} (\bibinfo{year}{1978}).

\bibitem[{\citenamefont{Varotsos et~al.}(1978)\citenamefont{Varotsos, Ludwig,
  and Falter}}]{ref25}
\bibinfo{author}{\bibfnamefont{P.}~\bibnamefont{Varotsos}},
  \bibinfo{author}{\bibfnamefont{W.}~\bibnamefont{Ludwig}}, \bibnamefont{and}
  \bibinfo{author}{\bibfnamefont{C.}~\bibnamefont{Falter}},
  \bibinfo{journal}{Journal of Physics C: Solid State Physics}
  \textbf{\bibinfo{volume}{11}}, \bibinfo{pages}{L311} (\bibinfo{year}{1978}).

\end{thebibliography}

\end{document}